# HIDDEN TIME AT WORK:
## SIMPLE INTERFERENCE, DELAYED CHOICE, ENTANGLEMENT

Pavel V. Kurakin

Keldysh Institute of Applied Mathematics, Russian Academy of Sciences
Mailto: kurakin.pavel@gmail.com

Previously we suggested [1 – 3] a concept of hidden time for building dynamical model underlying quantum phenomena. This paper brings a detailed hidden time explanation of some basic quantum experiments.

> "The problem is that the theory is too strong, too compelling. I feel we are missing a basic point. The next generation, as soon as they will have found that point, will knock on their heads and say: How could they have missed that?"
>
> I. I. Rabi.

## I.  Simple interference

**1**

Let us assume most fundamental quantum phenomenon: the interference. Fig.1 shows Mach – Zehnder interferometer. Experimenter supplies one photon in direction *1*. Standard quantum – mechanical description of this experiment is as follows.

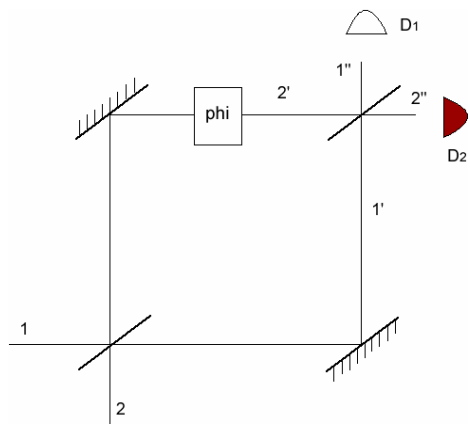

*Fig.1*
Simple 1 – photon interference in Mach – Zender interferometer

At the 1$^{st}$ 50\50 beam splitter quantum state of the photon is split in the superposition of passed and reflected photons:

$$|\psi_1\rangle \to \frac{1}{\sqrt{2}} \cdot (|\psi_{1'}\rangle + i \cdot |\psi_{2'}\rangle).$$

Recall that beam splitters rotate the phase of a photon state by $\frac{\pi}{2}$. Next, each mirror also rotates each state in superposition (multiplies by *i*):

$$|\psi_{1'}\rangle + i \cdot |\psi_{2'}\rangle \rightarrow i \cdot (|\psi_{1'}\rangle + i \cdot |\psi_{2'}\rangle).$$

Next, *2'* path state is rotated by φ, and before falling onto 2<sup>nd</sup> 50\50 beam splitter, the photon's state is:

$$|\psi_1\rangle \rightarrow \frac{i}{\sqrt{2}} \cdot (|\psi_{1'}\rangle + i \cdot \exp(i \cdot \varphi) \cdot |\psi_{2'}\rangle) \quad (1).$$

After passing the 2<sup>nd</sup> beam splitter, the state *1'* is split into superposition of states *1''* and *2''*:

$$|\psi_{1'}\rangle \rightarrow \frac{1}{\sqrt{2}} \cdot (|\psi_{1''}\rangle + i \cdot |\psi_{2''}\rangle) \quad (2).$$

The same does the *2'* state:

$$|\psi_{2'}\rangle \rightarrow \frac{1}{\sqrt{2}} \cdot (|\psi_{2''}\rangle + i \cdot |\psi_{1''}\rangle) \quad (3).$$

Summing (2) and (3) and inserting in (1), finally we get:

$$|\psi_1\rangle \rightarrow \frac{i}{2} \cdot [(1 - \exp(i\varphi)) \cdot \psi_{1''} + i \cdot (1 + \exp(i\varphi)) \cdot \psi_{2''}] \quad (4).$$

As a result the probability for detector D₁ to detect the photon:

$$P_1 = \left|\frac{1}{2} \cdot (1 - \exp(i\varphi)) \cdot |\psi_{1''}\rangle\right|^2 = \frac{1}{4} \cdot |1 - \exp(i\varphi)|^2 = \sin^2\left(\frac{\varphi}{2}\right) \quad (5\text{-}1).$$

Accordingly, the probability for detector D₂ to detect the photon is $P_2 = \cos^2\left(\frac{\varphi}{2}\right)$ (5-2).

If we have no phase shifter at path *2'* at all (φ = 0), we get that detector D₁ never clicks. Instead, detector D₂ clicks each time.

**2**

Here's what hidden time concept provides for this experiment.

1. At first, *scout* waves propagate from 1<sup>st</sup> beam splitter to detectors by all possible paths (fig. 2a).

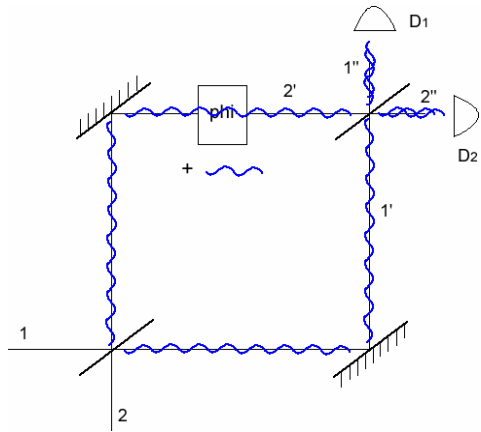

*Fig.2a*
Hidden time at work: *scout* waves come to each detector by all paths

Scout waves propagate like standard classical electromagnetic waves. This is why we put two mutually shifted waves at paths *1''* and *2''*: they originate from summing *1'* and *2'*. Please note that phase shifter provides some additional rotation of wave phase in the *2'* path.

Note also that physical time *does not go* while these waves "move".

2. When any detector receives a scout wave, it sends a *query* wave (Fig. 2b). Query wave *can* be just reversed in "time" variable scout wave, since Maxwell equations for free electromagnetic field is time – reversible. But in fact only *intensity* $|\psi|^2$ of query waves matters, not the phase. Intensity is a measure of *survivability* of waves sent by different detectors.

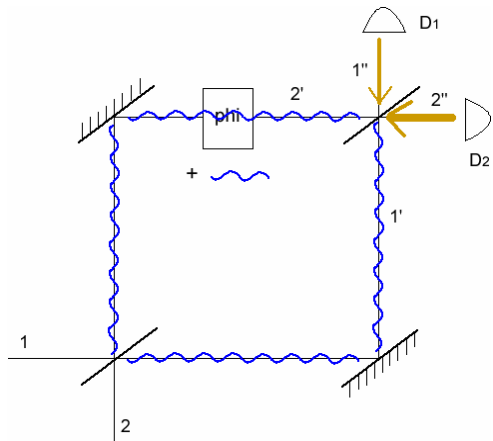

*Fig.2b*
Hidden time at work: *query* waves start to compete

At Fig. 2b a *query* wave from $D_2$ is stronger (say, φ in (4) and (5) is rather small, so $P_2 > P_1$), so it has more chances to win a *query* wave from $D_1$. Still, as one can see from Fig. 2c, we admit that nevertheless $D_1$ wins. I do so to express that intensity of a query wave means only the *probability* to win. $D_2$ has, generally speaking, a greater probability to win, but this time $D_1$ wins. If we repeat the experiment many times, the number of each detector's clicks is proportional to its *query* wave intensity $|\psi|^2$. It is exactly what standard quantum mechanics provides.

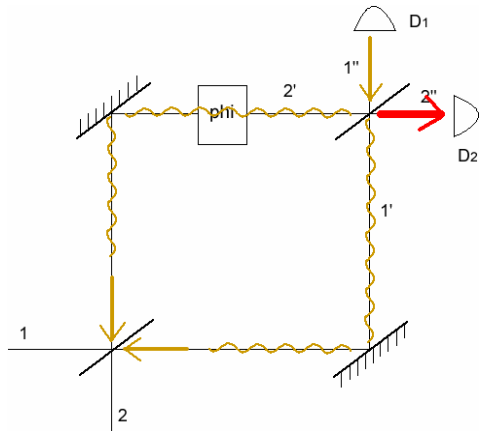

*Fig.2c*
Hidden time at work: the winner *query* wave propagates

After winning at $2^{nd}$ beam splitter, *query* wave of $D_1$ propagates next by all possible ways, i.e. by ways provided by *scout* waves before (Fig. 2c), i.e. by *1'* and *2'*. Each way's copy of that wave comes to $1^{st}$ beam splitter and they also compete there. Since they are of equal intensity, they have equal chances to win the competition. In fact, it is not essential which one wins, because each of these waves is representative of the same single detector $D_1$.

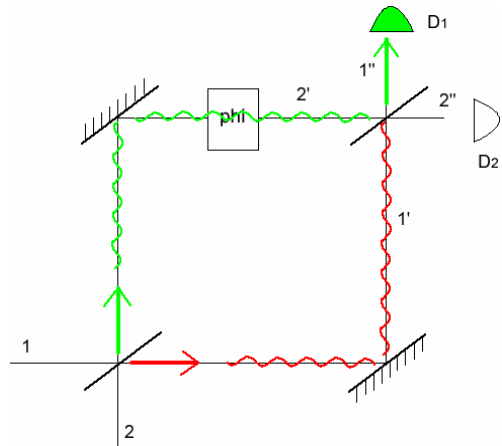

*Fig.2d*
Hidden time at work: detector $D_1$ gets the photon

At Fig. 2d *2'* path wins. Green signal means propagation of *confirmation* wave, which finally comes to detector $D_1$. Red signal is refusal wave (just like at Fig. 2c), it propagates up to it's source, i.e. $2^{nd}$ beam splitter.

**3**

The trick here is that all the waves represented above *do not* propagate in physical time. Only coming of confirmation wave (green) to $D_1$ corresponds to an instant of physical time. Though we have 3 passes of waves from a source to some detector, and though *each* of these passes obeys Maxwell's equations for free electromagnetic field, finally total time for a photon to travel to the detector is such as if the photon simply moves *classically* from the source to the detector with light speed!

This sounds very strange, but it is correct if one takes 4 postulates:

**(a)** *Any* photon's exchange by charges is accomplished through presented mechanism of 3 passes of waves;
**(b)** *Only* final coming of *confirmation* wave to some detector (a charge) corresponds to an instant of physical time;
**(c)** Each detector charge serves all scout waves from *different* source charges it receives *consequently*; this means that a detector charge sends a query wave in response to some scout wave only after a confirmation (or a refusal) is received for a previously sent query wave;
**(d)** *Detected* physical time is simply a *number* of *detected* light quanta, properly normalized for each experimental configuration.

One can find more detailed arguments in [2] and [3].

## II.   Delayed Choice

**1**

"Delayed choice" experiment was originally suggested by J. A. Wheeler as a *gedankenexperiment*. At the moment many versions of delayed choice experiment are proposed and realized. We assume the simplest version as it is presented in [4].

Let us take that we have no phase shifter at the path *2'* of our Mach – Zehnder interferometer at Fig. 1: $\varphi = 0$. In this case, $D_1$ never clicks; instead, $D_2$ clicks each time.

Next, suppose that "at last moment" experimenter decides to perform another experiment than he originally intended: he pulls $2^{nd}$ beam splitter out of the photon's path *2'*. "At the last moment" means, of course, before a moment of time, when classically moving photon *should* reach the beam splitter.

In this case, the photon hits $D_1$ and $D_2$ with equal probabilities. In other words, pulling out $2^{nd}$ beam splitter "at the photon's nose" gives the same result as if there were no $2^{nd}$ beam splitter from the very beginning.

From standard quantum mechanical viewpoint, the explanation is as follows. We take into account *all* classically possible paths in space-time for the photon; each path provides its own complex-valued amplitude $\psi_i$. Total amplitude of considered transition is the sum $\sum \psi_i$ for all paths.

So, when the $2^{nd}$ (upper at Fig. 1) beam splitter is removed from the interferometer at a proper time, each classically possible path in space-time *does not* encounter that beam splitter on its way. Thus destructive interference for $D_1$ disappears.

**2**

What explanation will hidden time approach provide for delayed choice experiment? First, let us take for simplicity that $2^{nd}$ beam splitter is not removed indeed from it its position; instead it is in some way switched off to be transparent for light. Switching off is accomplished by some electromagnetic signal. Again, we can take for simplicity that that this signal consist of a single quantum of light.

Recall that according to rule (a) (Section 3) this quantum propagates obeying the same general mechanism we outlined before. This means that initially a *scout* wave of this quantum arrives to the beam splitter (Fig. 3a). We assume that we switch off the beam splitter "at the nose" of the photon which propagates within the interferometer. This means that *scout* wave of switching photon arrives (in *hidden* time!) to the beam splitter earlier than the *scout* wave of interferometer photon.

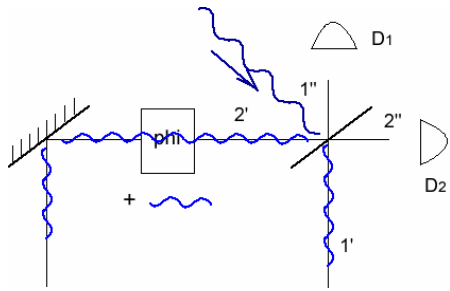

*Fig.3a*
Hidden time at work: *scout* wave of control photon comes to 2$^{nd}$ beam splitter

According to rule (c) from Section 3, this means that the beam splitter (we take it as if it is a single charge) serves switching photon first. Interferometer's photon scout waves stands unserved (and with no further propagation) until switching photon is ultimately served: i.e. until it is absorbed or rejected by the beam splitter.

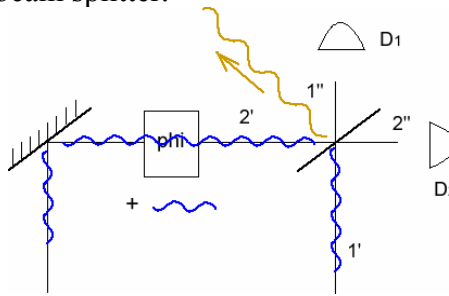

*Fig.3b*
Hidden time at work: *query* wave of control photon comes from 2$^{nd}$ beam splitter

So, the beam splitter sends *query* wave in response to switching photon's *scout* wave (Fig. 3b). Then, it receives *confirmation* wave, i.e. the switching photon itself (Fig. 3c). Of course, we suggest that the beam splitter receives confirmation rather then refuse, since we assume that we *do* switch off the beam splitter!

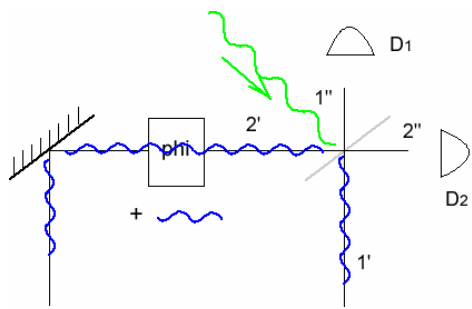

*Fig.3c*
Hidden time at work: *confirmation* wave of control photon comes to 2$^{nd}$ beam splitter

Only after switching off the beam splitter (which is equivalent to receiving confirmation wave in hidden time) *scout* waves of interferometer's photon propagate next from the beam splitter to detectors. And now they **do not** interfere (Fig. 3d), since the beam splitter is practically absent!

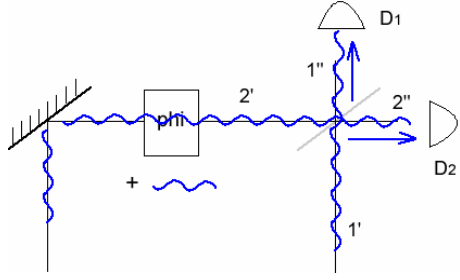

*Fig.3d*
Hidden time at work: scout waves don't interfere since beam splitter is transparent

So in this case *query* waves from $D_1$ and $D_2$ compete to each other at the 1$^{st}$ beam splitter, kike at Fig. 2d, and finally any detector can click.

### III. Entanglement: the core

Again, we start from what standard quantum theory provides for entanglement. In quantum optics domain we can get entangled photons most simply by 50\50 beam splitter (*Fig. 4*). Let us take that we send horizontally polarized photon to input 1 and vertically polarized photon to input 2 of the beam splitter. Since the 50\50 beam splitter adds a phase factor *i* to redirected "half" of each photon, we at the outputs:

$$|H\rangle_1 |V\rangle_2 \to \frac{1}{2} \cdot (|H\rangle_{1'} + i \cdot |H\rangle_{2'}) \cdot (|V\rangle_{2'} + i \cdot |V\rangle_{1'}) =$$
$$\frac{1}{2} \cdot (|H\rangle_{1'}|V\rangle_{2'} - |H\rangle_{2'}|V\rangle_{1'} + i \cdot (|H\rangle_{1'}|V\rangle_{1'} + |H\rangle_{2'}|V\rangle_{2'})) \quad (6).$$

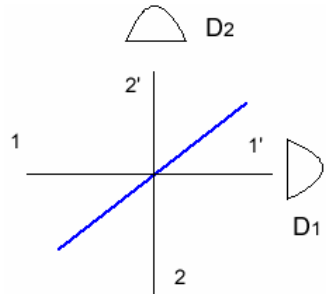

*Fig. 4*
50 \ 50 beam splitter produces entanglement of two photons

Under the condition of detecting a single photon per each detector, this state is finally projected onto

$$|H\rangle_1 |V\rangle_2 \to \frac{1}{2} \cdot (|H\rangle_{1'}|V\rangle_{2'} - |H\rangle_{2'}|V\rangle_{1'}).$$

This state is "maximally entangled": detecting $|H\rangle$ in one detector inevitably leads to detecting $|V\rangle$ in the other, and otherwise.

On the other side, imagine that we use not $|H\rangle$ and $|V\rangle$ polarizations for input photons; instead, we use arbitrary $|P_1\rangle$ and $|P_2\rangle$ – the derivation of final result does not depend on this substitution. Next, let us take that $P_1 = P_2 = P$, i.e. our photons are indistinguishable. In that case, (6) reduces to

$$|P\rangle_1 |P\rangle_2 \rightarrow \frac{i}{2} \cdot \left( |P\rangle_{1'}|P\rangle_{1'} - |P\rangle_{2'}|P\rangle_{2'} \right) \qquad (7).$$

State (7) means that each time only one detector clicks or, in other words, the photons always go together and never separately! It is Hong – Ou – Mandel effect.

**2**

Here's what hidden time concept can offer to explain entanglement. At the inputs of 50\50 beam splitter we have $|H\rangle$ *scout* wave at input *1* and $|V\rangle$ wave at input *2* (Fig. 5a).

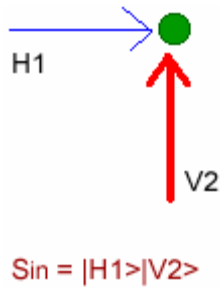

*Fig 5a*
Entanglement: the input state

Recall that the beam splitter splits each wave:

$$|\psi_1\rangle \rightarrow \frac{1}{\sqrt{2}} \cdot \left( |\psi_{1'}\rangle + i \cdot |\psi_{2'}\rangle \right).$$

Recall also that in hidden time model each term of this sum represents some new *thread* of a *scout* wave. When we multiply such sums, from hidden time view, this means that each term thread in each sum becomes "tied" or "married" to each term thread in the other sum. So, each term in each factor is split again, while the factor of splitting is the number of terms in paired factor.

Thus after leaving the beam splitter we have 2 × 2 = 4 *groups* of *tied* (or *married*) scout waves (Fig. 5b):

$$S^1_{out} = |H_{1'}\rangle |V_{2'}\rangle,$$
$$S^2_{out} = |H_{1'}\rangle |V_{1'}\rangle,$$
$$S^3_{out} = |H_{2'}\rangle |V_{1'}\rangle,$$
$$S^1_{out} = |H_{2'}\rangle |V_{2'}\rangle.$$

Let us refer to each group of tied threads as a *bunch*. Each of these bunches of scout waves correspond (with appropriate factors) to some term in (6) (when in expanded view!). Each bunch has a kind of *connection knot*, where single – photon waves are "tied". Any scout wave within bunch can be viewed as to be "married" with its partner. We mark it by a kind of "marriage rings" at Fig. 5b.

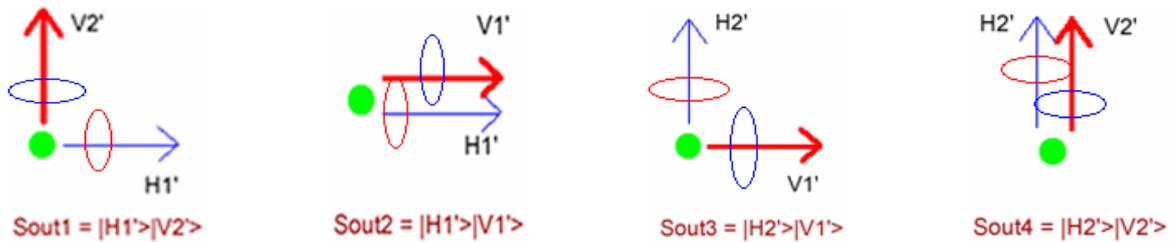

*Fig 5b*
Entanglement: 4 bunched output states

When we say "*the second photon*" we mean that it is in some way *distinguishable* from the first photon. Turn back to a series of Fig 2. The second beam splitter of the Mach – Zehnder interferometer *does not* make *additional* splitting of daughter scout waves, since they are *indistinguishable*!

In standard quantum mechanical formalism we **add** amplitudes for *indistinguishable* events and **multiply** them for *distinguishable* (independent) events. The same we do for scout wave *threads* in hidden time theory!

**3**

Next, imagine that each one – photon *scout* wave in each *bunch* behaves just the same we described above: it travels to appropriate detector and is (possibly) summed there with identical copies of itself (if any)[1], like it takes place in simple interference. Then that detector sends a *query* wave with intensity equal to squared sum of received *scout* waves.

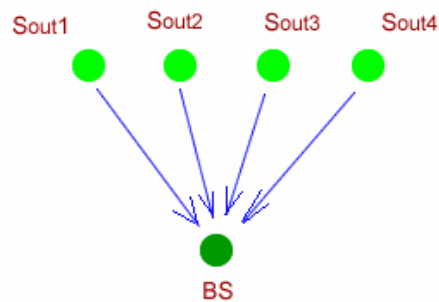

*Fig 5c*
Output states compete as independent units

Previously we suggested that the phase is inessential for a *query* wave. Let us now assume that it is not so: a *query* wave has phase variable, but it is constant along the way back from appropriate detector, and equal to the phase of scout wave at this detector. So each (one – photon) query wave provides a <u>standard (one - photon!) quantum mechanical</u> amplitude for a particular detector.

When any (one – photon) *query* wave comes from its detector to *bunch connection knot* (green circle in Fig. 5b) it meets its partner within the bunch. Amplitudes of both waves are *multiplied* as complex numbers. So we get each of the values of $S_{out}^1$, $S_{out}^2$, $S_{out}^3$, $S_{out}^4$ to be physically calculated in hidden time.

---

[1] That should be copies within the same *bunched* state! In other words, each one – photon scout wave within a bunched thread can divide and converge along its way like in 1 – photon Mach – Zehnder interferometer.

The following is quite trivial: bunched states compete to each other just like we discussed above for ordinary *query* waves in one – photon experiment (Fig. 5c). This competition leads for some *single* bunched state to win, and each detector gets its photon (if any) for this winner state.

### IV. Entanglement: change of basis

I want to argue that hidden time model of entanglement does not simply *looks like* true quantum entanglement, but is *exactly* equivalent to what quantum mechanics provides. Let we have an entangled two – component system:

$$\Psi_{AB} = \frac{1}{\sqrt{2}} \cdot (|0_A\rangle|1_B\rangle - |1_A\rangle|0_B\rangle) \quad (8).$$

Let we change observation basis in the following way:

$$|X_{A,B}\rangle = \frac{1}{\sqrt{2}}(|0_{A,B}\rangle + |1_{A,B}\rangle)$$
$$|Y_{A,B}\rangle_{A,B} = \frac{1}{\sqrt{2}}(|0_{A,B}\rangle - |1_{A,B}\rangle) \quad (9).$$

If $|0\rangle$ and $|1\rangle$ denote polarization states, then such change of basis means that we rotate optical axes of crystals we use by $\frac{\pi}{2}$ counterclockwise. From these equations we find that

$$|0_{A,B}\rangle = \frac{1}{\sqrt{2}}(|X_{A,B}\rangle + |Y_{A,B}\rangle),$$
$$|1_{A,B}\rangle = \frac{1}{\sqrt{2}}(|X_{A,B}\rangle - |Y_{A,B}\rangle). \quad (10).$$

If we substitute these expressions of into (8) we get:

$$\Psi_{A,B} = \frac{1}{2\sqrt{2}}[(|X_A\rangle + |Y_A\rangle)(|X_B\rangle - |Y_B\rangle) - (|X_A\rangle - |Y_A\rangle)(|X_B\rangle + |Y_B\rangle)] =$$
$$\frac{1}{2\sqrt{2}}[|X_A\rangle|X_B\rangle - |X_A\rangle|Y_B\rangle + |Y_A\rangle|X_B\rangle - |Y_A\rangle|Y_B\rangle - |X_A\rangle|X_B\rangle - |X_A\rangle|Y_B\rangle + |Y_A\rangle|X_B\rangle + |Y_A\rangle|Y_B\rangle]$$
$$= -\frac{1}{\sqrt{2}}(|X_A\rangle|Y_B\rangle - |Y_A\rangle|X_B\rangle)$$

So in (*X, Y*) basis we get *the same* entangled state (change of overall phase by $\pi$ has no physical sense) as in (*0, 1*) basis. It is crucial for us here whether hidden time model is able to reproduce this feature of standard quantum theory.

I suggest that such change of basis makes no difficulties for hidden time approach. Since at the input and output of a beam splitter we have $|0\rangle$ and $|1\rangle$ ket states only, we have *no any changes* of hidden time signals until they reach optical devices (birefringent crystals). In other words, in hidden time we have $|H\rangle|V\rangle$ single – photon waves and the same bunched waves for two photons as before:

$$S_{out}^1 = |H_{1'}\rangle|V_{2'}\rangle,$$
$$S_{out}^2 = |H_{1'}\rangle|V_{1'}\rangle,$$
$$S_{out}^3 = |H_{2'}\rangle|V_{1'}\rangle,$$
$$S_{out}^1 = |H_{2'}\rangle|V_{2'}\rangle.$$

Next, if we rotate the crystal located at the *1'* path, it is exactly this location, after which $|H_{1'}\rangle$ or $|V_{1'}\rangle$ scout wave (alone or within a bunch) is split into sum of ordinary $|o\rangle$ and extraordinary $|e\rangle$ waves – *Fig. 6*.

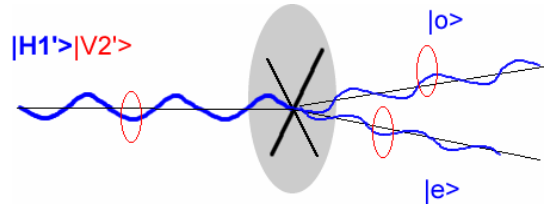

*Fig. 6*
Married |H1'> scout wave is split by birefringent crystal into married |o> and |e> scout waves, each of half intensity

Like before, $|o\rangle$ and $|e\rangle$ scout waves propagate to detectors, then come back to splitting point, i.e. the birefringent crystal we use, and compete there. Next, regardless of which of two branches $|o\rangle$ and $|e\rangle$ wins, initial scout wave $S_{out}^1 = |H_{1'}\rangle|V_{2'}\rangle$ propagates backward with its *own* intensity and phase, to the beam splitter we use to compete there with other *bunched* (married) states we have in the system.

In other words, from hidden time viewpoint, change of basis at detection location does not affect competition of initial scout waves.

So, within hidden time approach, entanglement is indeed *independent* of change of basis.

## V.     Entanglement: delayed choice

From previous description it should be clear that using time – varying analyzers in entanglement experiments, like in classical experimental work [5], brings *no new challenge* to hidden time approach.

The core idea is very simple:

- 1$^{st}$, scout waves find all detectors and "frozen" physical states of all devices involved, the very time is "frozen";
- 2$^{nd}$, scout waves come back to their sources and compete their;
- 3$^{rd}$, winner waves return to appropriate detectors and "unfrozen the time".

This general scheme just ignores our naive tricks with fast switching of analyzers. We think we are fast to switch the devices we use, but hidden does not think so.

## VI. Conclusions

It is notable that the idea of back propagation was formulated not once to interpret quantum mechanics [6, 7]. Still, the idea of hidden time, which a trivial and obvious *next* step, was not yet formulated.

I believe that hidden time can be much more than simply an *interpretation* of quantum mechanics; instead, it pretends to be the *underlying dynamics*. Why? Because hidden time model introduces some new and unexpected functions for charges: hidden time says something on **how a charge works with light in detailed algorithm**.